# Cancer Genesis and Progression as Dynamics in Functional Landscape of Endogenous Molecular-Cellular Network

Ping Ao[1], David Galas[2], Leroy Hood[2], Xiaomei Zhu[3]

[1] Department of Mechanical Engineering and Department of Physics, University of Washington, Seattle, WA 98195, USA
[2] Institute for Systems Biology, 1441 N. 34 St., Seattle, WA 98103, USA
[3] GenMath, Corp. 5525 27th Ave.N.E., Seattle, WA 98105, USA

March 23; September 4, 2007

**Summary**  An endogenous molecular-cellular network for both normal and abnormal functions is assumed to exist.  This endogenous network forms a nonlinear stochastic dynamical system, with many stable attractors in its functional landscape. Normal or abnormal robust states can be decided by this network in a manner similar to the neural network.  In this context cancer is hypothesized as one of its robust intrinsic states.

This hypothesis implies that a nonlinear stochastic mathematical cancer model is constructible based on available experimental data and its quantitative prediction is directly testable.  Within such model the genesis and progression of cancer may be viewed as stochastic transitions between different attractors. Thus it further suggests that progressions are not arbitrary. Other important issues on cancer, such as genetic *vs* epigenetics, double-edge effect, dormancy, are discussed in the light of present hypothesis. A different set of strategies for cancer prevention, cure, and care, is therefore suggested.

**Introduction**

Cancer as an enigma has been with humanity since the time unmemorable. It was recorded very early on in our written history [1]. In an 1806 issue of the Edinburgh Medical and Surgical Journal, 13 insightful and surprisingly modern-looking questions on cancer had already been formulated [2]. Since then there has been a tremendous progress in cancer research, particularly in elucidating its molecular, cellular, and epidemical connections during past 40 years. This exponential growth of molecular and cellular knowledge is aptly demonstrated by two classical cancer monographs, a two-volume set around 1970 [3] and a marvelous informative and updated one in 2006 [4]. Many facets of cancer can now be addressed with confidence. Nevertheless, the essential part of those 13 questions still remains open. In order to motivate a synthetic understanding of cancer, 7 questions have been recently formulated in the form of dichotomy [5], which clearly summarized the current state of art.

Despite such a long and bewildering history on cancer, it appears to us that the time to squarely face this disease computationally is coming. In the following we propose a hypothesis based on the existence of an endogenous molecular and cellular network. This endogenous network functions to cope with ever-changing living conditions: starvation, stress caused by change in habitants, sudden availability of abundant foods, infection, illness, wound, etc, shaped by hundreds of million year evolution of multi-cellular organisms [1]. Some of its essential pathways and modules may even be traced billions of years back to single cell organism stage. We will try to demonstrate that it can provide an overarch framework to understand cancer, to provide a guidance for its prevention, cure, and care. Through its validation, revision and extension, or through its falsification and possibly ultimate rejection, a better understanding and an eventual cancer theory may be reached.

**Hypothesis**

The hypothesis may be formulated as follows:
The molecular and cellular agents, such as oncogenes and suppressor genes, and related growth factors, hormones, cytokines, etc, form a nonlinear, stochastic, and collective dynamical network, the endogenous molecular-cellular network. This endogenous network may be specified by the expression or activity levels of a minimum set of endogenous agents, resulting in a high dimensional stochastic dynamical system. The nonlinear dynamical interactions among the endogenous agents can generate many locally stable states with obvious or non-obvious biological functions. The endogenous network may stay in any of such stable state for a considerably long time. In this manner the endogenous network is able to autonomously decide its operational functioning state. Some states may be normal, such as cell growth, apoptosis, arresting, etc. Others may be abnormal, such as growth with elevated immune response and high energy consumption, likely the signature of cancer, or of still useful functions to deal with occasional stressful situations. The stochasticity may accidentally cause a transition from one stable state to another. If with a given condition the endogenous network is in a state not optimized for the interest of whole organism, the organism is "sick", though this state might be "normal" under other conditions. Through the identifying agents of this endogenous network, the delineating of its wiring rules among endogenous agents, and the elucidating its global dynamical properties, a systems understanding of both normal and abnormal behaviors on how a tissue functions may be reached. In short, cancer is proposed as an intrinsic robust state of the endogenous network not optimized for the interest of whole organism.

There are 3 immediate and clarifying questions to be addressed.

1) Does such endogenous network exist?
Since the discovery of the first oncogene in 1970's [6], an impressive array of oncogenes and their suppressors have been discovered, covering almost all aspects of molecular and cellular functions. The existence of such a network has been repeatedly proposed and updated [7-9]. The recent monograph on cancer [4] again shows its existence. With this understanding, because those oncogenes and suppressor genes have normal functions and they are parts of endogenous molecular-cellular network, the terms such as "oncogenes" are somewhat misleading. They are among the key endogenous agents perhaps indispensable for a given tissue.

2) How does such a nonlinear stochastic dynamical network make its own decision?
Until we have a rigorous mathematical model, questions of this kind may always be with us. Nevertheless, we envisage that the general working of such endogenous network is similar to how the neural network of the brain works. At least two versions of nonlinear dynamical approaches for neural network have been proposed and have been seriously studied both experimentally and theoretically with varied degrees of success: one via a potential function [10] and other via chaotic attractors [11,12]. We further point out that the recent developments in mathematical tools on stochastic processes [13-17] provide a unified framework to make use of both approaches. The landscapes of multiple dimensional biological systems in stochastic environments have been explicitly studied recently, for example, for genetic switches [18-21], cell cycle [22,23], signal transduction pathways [24-26].

3) What would be the essential parts of the endogenous network?
We envisage that the oncogenes and other molecular and cellular agents first form pathways and modules. The pathways and modules then cross talk to each other to form the endogenous network. Such a hierarchical structure is similar to the modular organization principle discussed in Ref. [27]. The essential pathways and modules are illustrated in Fig. 1. We refrain here from drawing a detailed cross-talk wiring diagram, because the precise information on such autonomous endogenous network remains to be obtained from literature. Nevertheless, interesting readers may consult, for example, Fig.2 of Ref.[8] for an idea of such diagram along the input-output reasoning, where many essential endogenous agents were also listed, and the Fig. 9 of Ref.[9] which is more close to what envisaged here.

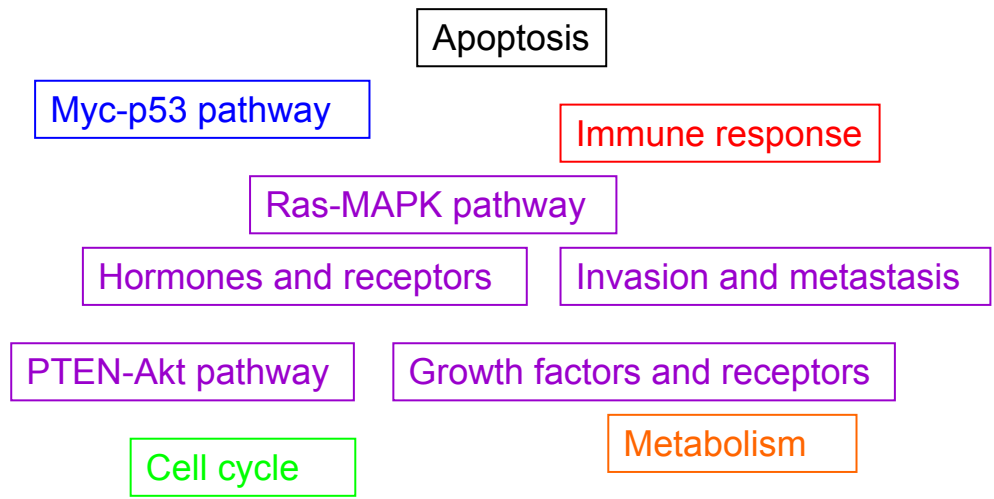

**Figure 1**.The minimum set of pathways and modules of the endogenous network. Endogenous molecular and cellular agents first form pathways and modules. Pathways and modules cross talk to each other to form the endogenous network.

Taking prostate cancer as a specific illustration, the molecular and cellular agents are, for example, E2F, pRb, and CyclinE/Cdk2 in cell cycle; PTEN and Akt in PTEN-Ark pathway; HIF in metabolism module; Androgen in hormone module; IGF-1 and VEGF in growth factor module; Myc and p53 in the Myc-p53 pathway; Ras and MAPK in Ras-MARK pathway; E-cadherin in invasion and metastasis pathway; NF-κB, TNF-α, and IL-10 for immune response module; and cytochrome c, Bad, caspace 3, Bcl-2, and XIAP for apoptosis module, already more than 20 endogenous agents. More can be easily listed, already discussed in Ref.[4].

**Implications**

It is clear that even above limited set of molecular and cellular agents in the endogenous network cannot be all produced within a single cell. This fact alone evidently suggests that the network properties can be affected both genetically, such as mutations, and epidemiologically, such as by other tissues. Therefore, the postulation of such endogenous network to understand cancer immediately provides a possible platform to understand the dichotomy between whether cancer is a genetic disease [28-32] or an epigenetic disease [33-37].

In the following we further discuss the compatibility of the proposed framework with a few pronounced observations on cancer based on a general and qualitative consideration of the hypothesis. We emphasize what will be demonstrated is plausibility, not precise and realistic quantitative explanations which have to be based on explicit mathematical models.

Let us begin with the possible number of locally stable states in the functional landscape of such endogenous network. As alluded above, there are at least 20 endogenous agents. It is known that mutual inhibiting or activating interactions are abundant in molecular and cellular pathways and modules, such E2F and CyclinE/Cdk2 [38] or cytochrome c and caspace 3 [39], a dynamical reciprocity recognized long ago [40]. Such a nonlinear and feedback interaction typically forms a bi-stable switch with 2 locally stable states. The classical example is the phage lambda genetic switch [41,18]. For a matter of demonstration, we may simply assume that there are, for example, 5 such independent effective dynamical reciprocal pairs formed by above 20 endogenous agents. Formation of such dynamical reciprocity pairs may take 10 endogenous agents. The activating-inhibiting pairs are also abundant in endogenous networks, such as MAPK----MAP [42] and iκB----NF-κB [43-44]. Their action may be regarded as buffering and regulation of those dynamical reciprocity pairs. Each activating-inhibiting pair normally forms one stable state. We may assume that there are effectively 5 of them, perhaps taking another 10 endogenous agents. The existence of many such elementary regulatory circuits has been demonstrated to be abundant in biological networks [45-47]. Then based on above hypothetic estimation of dynamical reciprocity and activating-inhibiting pairs, the total number of stable states, or the attractive basins in the functional landscape, may be 2x2x2x2x2x1x1x1x1x1 = $2^5$ x $1^5$ = 32. This number is by no means what should be in a real endogenous network, because how the endogenous agents are actually wired together would strongly affect this number. Nevertheless, we take this number as an anchor to indicate the possible number of stable states in the functional landscape of the endogenous network. For cases to be discussed below the conclusions are largely independent of the exact number of stable states.

Given the existence of multiple stable states, we may speculate that there are some states may correspond to healthy states under normal conditions, some to deal with rare stressful situations, and a few others would be the "disease" states or abnormal states. A schematic illustration of the functional landscape along a route going through minima and saddle points is in Fig. 2. Here it may be worthwhile to mention that the landscape idea has already explored early on in biology in other contexts: for example, the adaptive landscape in population genetics [48] and the developmental landscape in developmental biology [49]. Such idea has already been considered by molecular biologists [50].

With the existence of such functional landscape with multiple local stable states, the dormancy of neoplasma may be easily explained [51-53]: The endogenous network is in a state close to, but not in, the cancerous state. It shares many features of cancer. Because of the local stability as illustrated in Fig. 2, it can stay there for a very long time, thus the dormancy. Such "dormancy", or lysogenic state, has already been well known for a simple organism, phage lambda. Even after mutations in its genome, it may still stay in the lysogenic state for many generations [41,54]. Thus, the existence of functional landscape with multiple stable states explains the robustness widely observed in biological systems [18,24,55-60]. The existence of such multiple stable states may also corroborate well with the observations that cancer takes multiple steps [61]: If regarding moving from one stable state to another as one step.

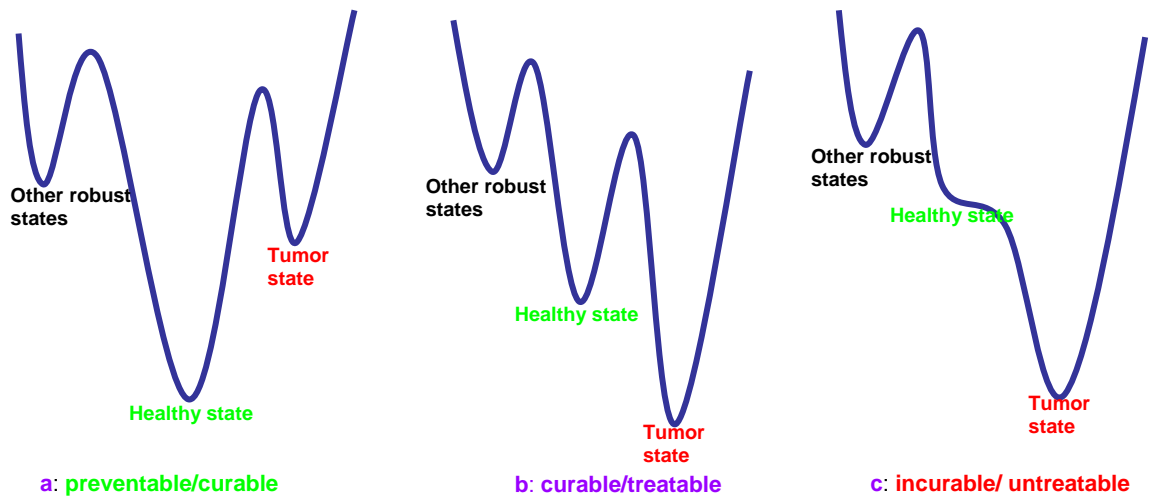

Figure 2. Three typical situations of the functional landscape. The vertical scale illustrates the relative stability of robust states, healthy, tumor and others, in the multiple dimensional state space. a) The healthy state is a globally stable under normal conditions; b) Due to genetic and epidemiologic influence on the endogenous network, tumor or cancer states may become more stable than healthy state. Such metastable healthy state may still have a long life time for the whole organism being viable; c) A very "damaged" endogenous network may not be able to produce a locally stable healthy state.

The existence of stable states in the functional landscape further suggests that the navigation from one stable state to another may not be arbitrary. There are preferred routes, most likely those passing through saddle points. This natural and qualitative consequence of the present mathematical consideration implies that the genesis and progression of cancer may have patterns. If this is true, present modeling methodology may suggest an alternative set of strategies for cancer prevention, cure, and care. It should be pointed out that such behaviors have already been noticed clinically [62]. Though the computational effort on such multiple-dimensional functional landscape may be heavy, it is an effort worthy pursuing.

The fact that the endogenous network can be influenced by both genetic and epidemiologic factors, as illustrated by three typical situations in Fig. 2 and discussed in its caption, implies that the network wiring can be changed, the network flexibility already noticed [63]. It certainly implies that other factors, such as chaperones, can also influence its functional landscape [64]. For example, certain level of chaperones can lower the barrier and make the local minimum shallower, resulting in larger fluctuations. Nevertheless, we believe the backbone and essential structure of the endogenous network would remain the same and is conserved, because it has been shaped by millions, or even billions, years of evolution. During the lifetime of an organism, there is a little chance of any major modification of the essential structure of the endogenous network.

Because interactions among endogenous agents are nonlinear, a few key endogenous agents may have self-amplifying effects. This self-amplifying may lead to a locally stable state in the functional landscape. Such the self-trapping effect is known in nonlinear dynamics. Those behaviors may correspond to the observed "oncogene addiction" [65].

It is known that the differentiation of stem cell is stochastic with multiple intermediate stable states [66-70]. The present functional landscape has the capacity to explain those observations: The navigating of endogenous network through those stable states in functional landscape may correspond to what observed. This ability of navigating through the functional landscape may suggest itself as a guiding principle to look for strategies of cure and care and well as prevention. Similar idea has been elaborated in literature [62], already discussed above.

Consider a possible extreme situation which appears critical important from basic biological consideration for present formulation, though it might not be clinically important. The existence of multiple states in functional landscape suggests that there can be a nonzero probability that cancer be formed without any genetic defect. In the same vein of reasoning, there can be a nonzero probability of cancer regression without any major interventions. We do not aware of any solid experimental evidence on those two suggestions, but it would be interesting to know how they would have been vindicated by experimental observation and clinical data.

We emphasize that there should be a minimum set of endogenous molecular and cellular agents in order to give a comprehensive description of the endogenous network. Such number is perhaps no less than 20, as indicated by above prostate cancer example. If one would only focus on one or a few endogenous agents, or one pathway and module, ambiguity may be resulted in. For example, cell proliferation or death can happen with either lower or higher immune activity. This may be related to the well known double-edge effect [71-72]. There exists another possibility: Because the dynamics is stochastic, a deterministic set of values for the endogenous agents may not provide us with the crucial information. The distributions and correlations among the endogenous agents should be measured experimentally or computed theoretically in order to give a better characterization of the endogenous network. There exists an even more intriguing possibility which may make the interpretation of experimental data more difficult. Because the endogenous agents are highly correlated to each other, a better description may be a set of collective variables of those endogenous agents. If such situation occurs, the information on the distributions and correlations of endogenous agents, if we do not have a good idea of what the collective variables were, may not provide an immediate characterization of the endogenous network. One needs to look for such collective variables. The present proposal may provide a mathematical and computational tool to address this last question, too.

Given all above considerations, it appears that the hypothesis of cancer as robust intrinsic state of the endogenous molecular-cellular network provides a framework to address all 7 dichotomies analyzed in Ref.[5].

**Conclusions**

We have proposed to understand cancer as robust intrinsic state of the endogenous molecular-cellular network in a manner similar to the neural dynamical modeling. A stochastic mathematical modeling scheme is implied in our proposal. We have demonstrated that there is enough experimental evidence to support the existence of such endogenous network. Discussions on immediate consequences of our proposals have indicated a qualitative agreement with all major observations on cancer. Here we further suggest that if this proposal is true for cancer, it may also be true for other complex diseases such as diabetes II. The immediate task is then to build from available experimental data a mathematically relative simple but biologically nontrivial quantitative model, to perform the analysis of such model as discussed above, to compare its predictions to experimental observations, to further improve the mathematical modeling. Through such interactive processes between experimental and theoretical efforts, a better understand of cancer should be reached.

**Acknowledgements:**
Critical and stimulating discussions with J.L. Abkowitz, L.A. Loeb, J. Roberts, S. Stolyar, particularly R.T. Prehn, are of great help. This work was supported in part by USA National Institutes of Health under HG002894 (P.A.).

*Note on literature:* Hundreds of billions dollars have already been spent on cancer world wide. The number of existing papers is measured in million. Therefore, no attempt had been made to provide a complete literature: We simply could be able to do that. Nevertheless, the authors believe that a good part of the current cancer research spirit has been captured, which leads to the network and evolution hypothesis presented in the text. In the light of such understanding, the authors will very much appreciate suggestions, comments and critiques on our omissions, ignorance, and misunderstandings (contact: aoping@u.washington.edu).

**Appendix. 13 questions and 7 dichotomies on cancer.**

Questions formulated 200 years ago are here. For a penetration discussion, see Ref.[2].
   "1) What are the diagnostic signs of cancer?
   2) Does any alteration in the structure of a part take place, preceding that more obvious change which is called cancer; and if there be an alteration, what is its nature?
   3) Is cancer always an original and primary disease; or may other diseases degenerate into cancer?
   4) Are there any proofs of cancer being an hereditary disease?
   5) Are there any proofs of cancer being a contagious disease?
   6) Is there any well-marked relation between cancer and other diseases? If there be, what are those diseases to which it bears the nearest resemblance in its origin, progress, and termination?
   7) May cancer be regarded at any period, or under any circumstances, merely as a local disease? Or does the existence of cancer in one part afford a presumption that there is a tendency to a similar morbid alteration in other parts of the animal system?
   8) Has climate or local situation any influence in rendering the human constitution more or less liable to cancer, under any form, or in any part?
   9) Is there any particular temperament of body more liable to be affected with cancer than others? If there be, what is the nature of that temperament?
   10) Are brute creatures subject to any disease resembling cancer in human body?
   11) Is there any period of life absolutely exempt from the attack of this disease?
   12) Are the lymphatic glands even affected primarily in this disease?
   13) Is cancer, under any circumstances, susceptible of a natural cure?"

The dichotomies are listed here for a quick comparison. Please check Sporn's beautiful elaborations [5].
   "i) 'The disease is cancer' versus 'the disease is really carcinogenesis';
   ii) 'Emphasis on cure of end-stage disease' versus 'prevention of early disease progression';
   iii) 'Cancer is a genetic disease' versus 'cancer is also an epigenetic disease';
   iv) 'New emphasis on transgenic mouse models with single gene disruption' versus 'classical carcinogenesis models that damage multiple genes';
   v) 'New emphasis on monofunctional agents' versus 'need for multifunctional agents'
   vi) 'Reductionism' versus 'the whole can be greater than the sum of its parts';
   vii) 'Hypothesis-driven research' versus 'the need for observational research'."

**Fig. 9 in [9] B. Vogelstein and K.W. Kinzler, 2004, Cancer genes and the pathways they control, Nature Medicine 10: 789-799.**

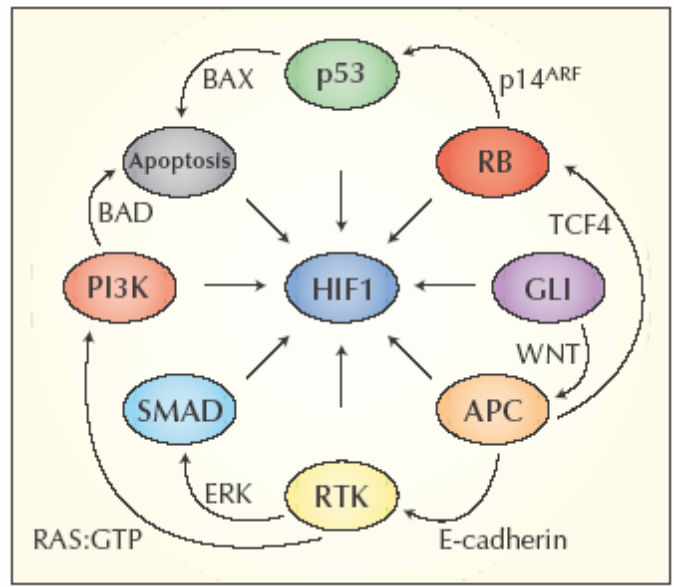

Figure 9 Overview of cancer gene pathways. The major pathways regulating cell birth and cell death are depicted as ovals color-coded to match Figs. 1–8. The schematics in Figs. 1–8 emphasize the genes that have been shown to be genetically altered in human tumors, though many other genes participate in these pathways. Additionally, some of the same genes appear in more than one pathway and there is substantial 'cross-talk' between pathways. Selected mediators of this cross-talk are indicated in the loops that connect the pathways. More detailed information about these pathways can be found in several comprehensive reviews (refs. 11,12,15,31,103–116).

**Fig. 2 in [8] D. Hanahan and R.A. Weinberg, 2000, The hallmarks of cancer, Cell 100: 57-70.**

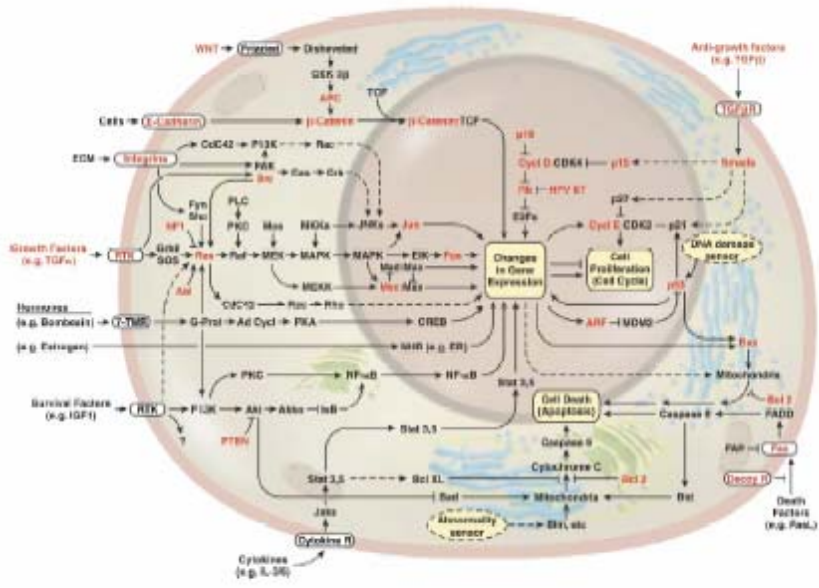

Figure 2. The Emergent Integrated Circuit of the Cell
Progress in dissecting signaling pathways has begun to lay out a circuitry that will likely mimic electronic integrated circuits in complexity and finesse, where transistors are replaced by proteins (e.g., kinases and phosphatases) and the electrons by phosphates and lipids, among others. In addition to the prototypical growth signaling circuit centered around Ras and coupled to a spectrum of extracellular cues, other component circuits transmit antigrowth and differentiation signals or mediate commands to live or die by apoptosis. As for the genetic reprogramming of this integrated circuit in cancer cells, some of the genes known to be functionally altered are highlighted in red.

**Fig. 2 in [7] T. Hunter, 1997, Oncoprotein networks, Cell 88: 333-346.**

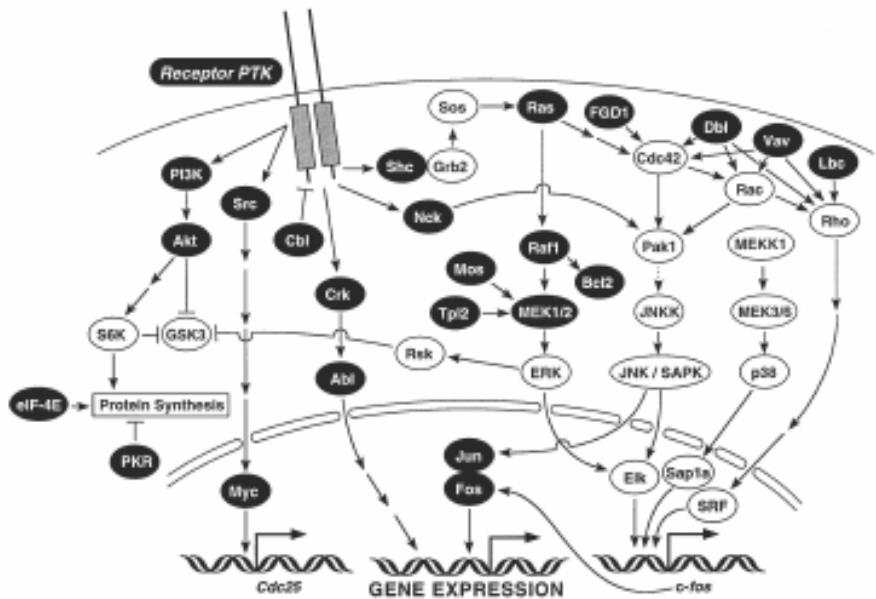

Figure 2. Membrane-to-Nucleus Signaling Pathways Involved in Cancer
Proteins that have been implicated as oncoproteins or tumor suppressor proteins are in white letters on black fill.

13